\documentclass[10pt,conference]{IEEEtran}
\IEEEoverridecommandlockouts
\usepackage{cite}
\usepackage{amsmath,amssymb,amsfonts}
\usepackage{algorithmic}
\usepackage{graphicx}
\usepackage{textcomp}
\usepackage{xcolor}
\usepackage{braket}
\usepackage{subcaption}
\usepackage{stfloats}
\usepackage{gensymb}

\begin{document}

\title{Interconnection of Quantum Networks \\at Urban scale: \\
Analysis of Temporal Stability \\of Entangled Photon Sources

\thanks{This work has been funded by the European Union under Horizon Europe ERC-CoG grant QNattyNet, n.101169850. Views and opinions expressed are however those of the author(s) only and do not necessarily reflect those of the European Union or the European Research Council Executive Agency. Neither the European Union nor the granting authority can be held responsible for them.}
}

\author{\IEEEauthorblockN{Laura d'Avossa}
~\IEEEmembership{Graduate~Student~Member,~IEEE}
\IEEEauthorblockA{\textit{University of Naples Federico II} \\
Naples, Italy \\
laura.davossa@unina.it}
\and
\IEEEauthorblockN{Angela Sara Cacciapuoti}
~\IEEEmembership{Senior~Member,~IEEE}
\IEEEauthorblockA{\textit{University of Naples Federico II} \\
Naples, Italy \\
angelasara.cacciapuoti@unina.it}
\and
\IEEEauthorblockN{Marcello Caleffi}
~\IEEEmembership{Senior~Member,~IEEE}
\IEEEauthorblockA{\textit{University of Naples Federico II} \\
Naples, Italy \\
marcello.caleffi@unina.it}
}

\maketitle

\begin{abstract}
Time synchronization is a fundamental requirement in entanglement-based quantum networks, where the indistinguishability of photons in the time domain is essential for enabling Hong–Ou–Mandel interference and entanglement swapping. In addition to precise temporal alignment, it is equally crucial to ensure the stability of the reference clock over time, as even small fluctuations can degrade the overall performance of the network.
In this work, we investigate the stability of clock synchronization for entanglement distribution based on entangled-photon sources operating in the telecommunication C-band.  
Temporal correlations between photon detection events are analyzed using time-tagged coincidence measurements, enabling the extraction of synchronization peaks and their long-term stability. Experimental results demonstrate that, once locked, the sources exhibit stable temporal correlations over an 8-hour acquisition period, with a maximum observed drift of approximately 120 ps, primarily associated with long fiber links. The width of the correlation peak remains consistent with detector jitter, indicating negligible additional system-induced temporal broadening.
A central result of this work is the experimental synchronization between two entanglement-photon sources in a realistic metropolitan deployment, where entanglement distribution is performed over existing telecommunication infrastructure characterized by non-negligible losses and background noise. In this scenario, despite the presence of significant imperfections introduced by the metropolitan-scale fiber network, the observed correlations remain clearly detectable and are consistently well-approximated by Gaussian statistics. This confirms that the two sources can be reliably synchronized not only in controlled laboratory conditions but also under real-world operating constraints, where channel impairments and environmental fluctuations play a dominant role.
Therefore, these results represent an initial step toward scalable quantum network architectures. 
\end{abstract}


\section{Introduction}
\label{sec:01}

\begin{figure*}[t]
\centering
\includegraphics[width=\textwidth]{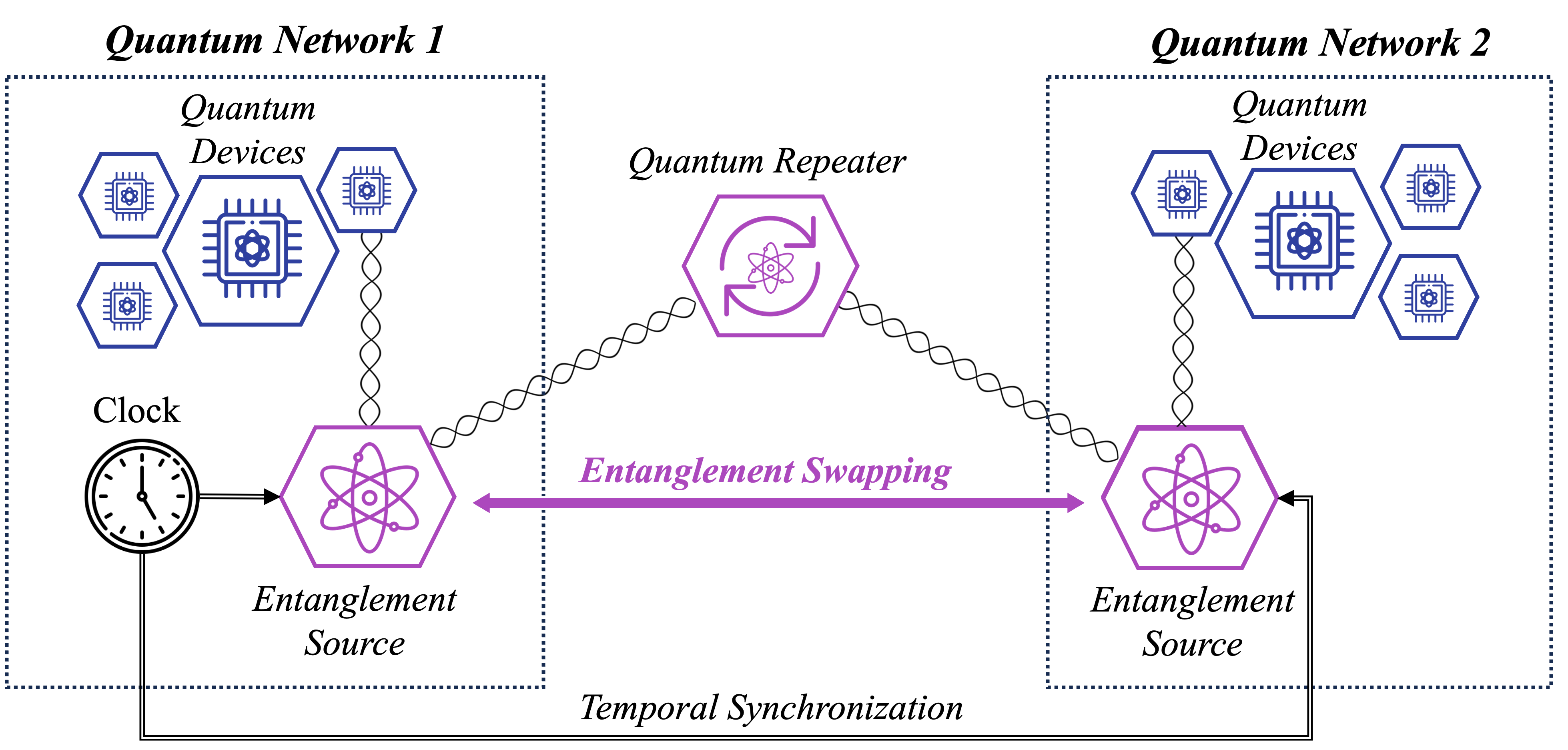}
\caption{Schematic representation of interconnected quantum networks. Multiple independent quantum networks are synchronized via a common reference clock. A quantum repeater node performs entanglement swapping between networks, enabling the realization of a large-scale quantum communication architecture.}
\hrulefill

\label{fig:01}
\end{figure*}

Entanglement is the key resource of the Quantum Internet, since it enables protocols with no counterpart in the classical domain \cite{CacCalTaf-19, CacCalVan-20}. 
The ability to distribute entangled states is essential for the development of scalable quantum networks and for enabling the implementation of complex quantum communication protocols.
For entanglement distribution, the synchronization of entangled photon sources is a fundamental requirement.
Entanglement swapping provides a representative example, as it critically relies on precise temporal synchronization to enable long-distance entanglement distribution, thereby extending the achievable communication range \cite{VanTou-13}.
This capability is the enabling principle for the interconnection of different quantum networks, giving rise to a broader, scalable quantum infrastructure. 
A schematic representation of this concept is shown in Figure~\ref{fig:01}, which shows two quantum networks interconnected through entanglement swapping enabled by synchronized entangled photon sources.
Entanglement swapping is performed in quantum repeater nodes \cite{AzuEcoElk-23}, which implement Bell state measurements (BSM) \cite{BrauMan-95}.
To perform BSM, photons from independent sources must be indistinguishable in all relevant degrees of freedom, including temporal, spectral, and polarization degrees of freedom. This indistinguishability is an essential condition for ensuring high-visibility quantum interference, in particular the Hong–Ou–Mandel (HOM) interference \cite{HonOuMan-87}, which constitutes the physical mechanism underlying BSM measurements based on two-photon interference.
Consequently, extremely precise \textit{temporal synchronization between the sources} is required to ensure the simultaneous arrival of photons at the interference nodes. Indeed, even minimal residual temporal misalignments can compromise the indistinguishability of the photons, significantly reducing the fidelity of the projected entangled states. This results in a direct degradation of the overall performance of the quantum networks.

Once the possibility of effectively distributing and controlling the reference clock among the sources has been verified, it becomes essential to analyze its stability over time—that is, its ability to maintain synchronization conditions within tolerances compatible with the requirements of quantum interference. This aspect is, in fact, crucial for the scalability and reliability of interconnection architectures between heterogeneous quantum networks. For this reason, synchronization that is not only precise but also stable over time represents an essential enabler for the development of truly scalable and robust quantum network architectures.

In this work, we experimentally demonstrate the distribution of a reference clock among different sources of entangled photons. The clock signal is fully controllable and can be dynamically modulated, both in terms of its period and the relative phase between the sources. We investigate the temporal stability of the internal synchronization mechanism of our entanglement source, with the aim of enabling reliable and robust entanglement distribution. In particular, we evaluate the performance of the internal synchronization mechanism in supporting entanglement distribution not only in a controlled laboratory environment, but also over a deployed metropolitan-scale fiber loop representative of pre-installed fiber infrastructure.

\section{Entanglement distribution via optical photons in the C-band}
\label{sec:02}

Among the different technologies for implementing entanglement carriers, optical photons have been widely recognized as the most promising candidates and therefore they are referred to as \textit{flying qubits}.
Indeed, they weakly interact with the environment, resulting in low decoherence, and they can be easily controlled with standard optical components. 
These properties make them optimal for long distance quantum communication \cite{CacCalTaf-19, CalDavHan-25, DavCacCal-25}.

The distribution of entanglement via optical photons has been widely investigated \cite{TalHesDav-26, TalHessJor-26}, with a key open question: determining which channel is the most suitable for entanglement distribution \cite{DavMonGri-26}.
The telecommunication C-band is considered an optimal choice for entanglement distribution, as it is also widely exploited for classical communication. Indeed, this portion of the spectrum, spanning the fourth DWDM channels\footnote{DWDM (Dense Wavelength Division Multiplexing) refers to a technology used in optical communications to multiplex multiple optical carrier signals onto a single fiber by using closely spaced wavelength channels, enabling high-capacity transmission and parallel quantum or classical communication channels \cite{Kar-02}.} $1520$-$1577$ nm, exhibits the lowest attenuation in standard single-mode fibers (SMFs) (approximately $~0.2$ dB/km \cite{ClaYanWanNejSimJos-24}) making it ideal for long-distance quantum communications.
This choice is also consistent with existing telecommunications infrastructure. Indeed, the large-scale deployment of quantum networks requires leveraging already installed fiber-optic systems, rather than relying on the installation of additional dedicated dark fibers. The latter would entail significant efforts and prohibitive costs, particularly in metropolitan environments, making such an approach impractical for widespread adoption.
In this scenario, the C-band proves to be the optimal choice for distributing entanglement over pre-installed fiber.
Accordingly, in this work we perform entanglement distribution exploiting C-band existing fiber infrastructure.

\begin{figure*}[t!]
\centering
\includegraphics[width=\textwidth]{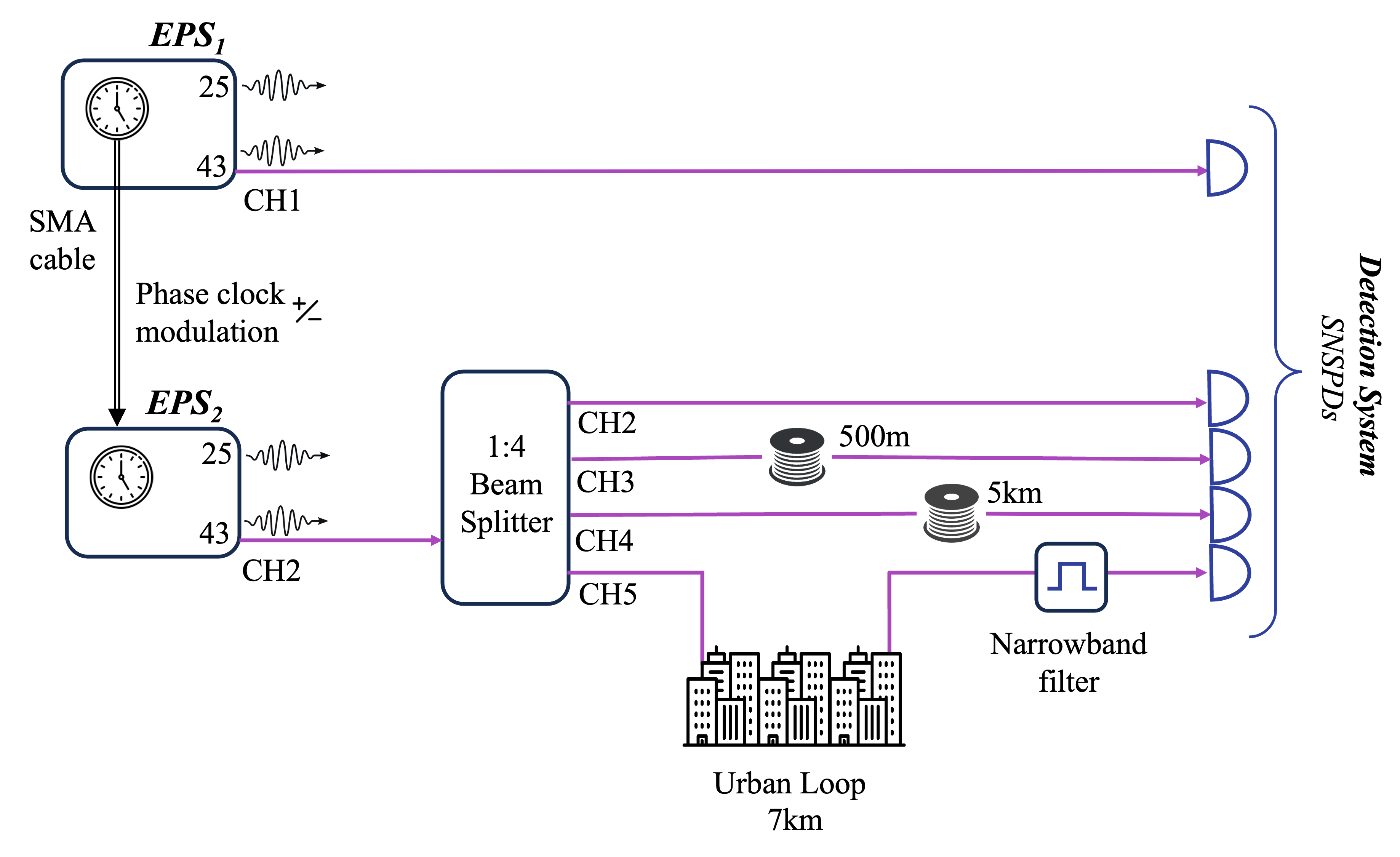}
\caption{Experimental setup built in the \textit{National Quantum Internet.it testbed} at the University of Naples Federico II. The setup consists of two polarization-entangled photon sources ($EPS_1$ and $EPS_2$), generating entangled pairs on DWDM channels 25-43. A shared electrical clock from $EPS_1$ is used to synchronize $EPS_2$. Photons from $EPS_1$ are directly sent to the detection stage, while photons from $EPS_2$ are distributed over multiple fiber links, including a short reference path ($\sim$10 m), a 500 m spool, a 5 km spool, and a 7 km metropolitan loop, in order to investigate timing stability under different propagation conditions. Detection is performed using SNSPDs, and arrival times are recorded via a time-tagging system.}
\hrulefill
\label{fig:02}
\end{figure*}

\section{Experimental setup}
\label{sec:03}

The analysis is conducted in the \textit{National Quantum Internet.it testbed} deployed at \textit{University of Naples Federico II}. 
The experiment involves two different campuses of the University. Part of the experiment is conducted at the Monte Sant'Angelo campus (MSA), where the laboratory equipped with entangled photon sources and a detection system is located; a second part of the experiment involves a metropolitan loop, which is described in detail in Sec-\ref{sec:05}.
A diagram of the system setup is depicted in Figure~\ref{fig:02}.
In the lab at the MSA campus, entangled photons are generated with two Entangled Photon Sources (EPSs). The sources are rack-mounted and plug-and-play devices able to generate ready-to-use fiber-coupled polarization entangled photons. 
Specifically, the EPSs generate polarization-entangled photon pairs through an internal Spontaneous Parametric Down-Conversion (SPDC). The generated state is namely: $\ket{\Phi^+}=\frac{1}{\sqrt{2}}(\ket{H_sH_i}+\ket{V_sV_i})$ where the subscripts $s$ and $i$ refer to the signal and idler, respectively. 

Each EPS supports multiple DWDM channels through an internal filtering system that allows flexible channel selection in the C band. In this work, channels 25 and 43 are chosen, as they are available through the source’s internal filtering, eliminating the need for external DWDM components.
In addition, each EPS provides access to an electrical clock with the same frequency of the internal pump pulses that drive the SPDC. This clock signal can be accessed via an external clock output pin. Each EPS can receive a reference clock via an electrical input pin.
To achieve synchronization between the two sources of entangled photons, the clock generated by one of the two sources, that we will call $EPS_1$ is used to control the second source via a SMA cable, that we call $EPS_2$, ensuring that the photons produced by the latter are thus synchronized with the same clock signal.
The entangled photons from both sources are sent through an SMF-28 fiber-optic network.

One photon from each entangled pair, namely, the signal photon generated by $EPS_1$ and the signal photon generated by $EPS_2$, both corresponding to ITU channel~43, is used in the experiment. The photon from $EPS_1$ reaches the detection system via a $\sim 10$~m long SMF-28 fiber installed in the MSA campus laboratory, while the corresponding photon from $EPS_2$ is sent through a 1:4 beam splitter.
The first output, serving as the reference path, also propagates through a $\sim 10$ m long fiber in the laboratory before reaching the detection system. The second, third, and fourth outputs propagate through a 500 m fiber spool, a 5 km fiber spool, and a deployed metropolitan fiber loop of approximately 7 km, respectively, before reaching the detection system.
At the return from the metropolitan loop, the channel 43 signal is filtered by a narrowband optical filter ($\approx 0.25$ nm bandwidth) to suppress spurious photons from other channels or noise accumulated along the link. 
For clarity each output port of the beam splitter is labelled in Figure~\ref{fig:02}. Namely, CH1 denotes the channel associated with $EPS_1$, while CH2–CH5 correspond to the different fiber links of $EPS_2$ ($\approx10$ m, 500 m, 5 km, and 7 km metropolitan loop, respectively).
The reason for choosing to use a beam splitter to split the traffic generated by $EPS_2$ is to study the stability of the clock shared between the sources over time across different fiber lengths, within the same time interval. Indeed, by splitting the $EPS_2$ traffic, it is possible to simultaneously analyze the effects of clock stability over multiple propagation lengths.
The detection system consists of a set of superconducting nanowire single-photon detectors (SNSPDs) optimized at 1550 nm\footnote{The SNSPDs have been fabricated, installed and tested by Photon Technology Italy in a rack mounted configuration \cite{photec2026}}. 
The measurements between the two sources are analyzed using a Swabian Ultra Performance time tagger with resolution of 5 ps.

\section{Clock Synchronization and Control}
\label{sec:04}

\begin{figure}[!t]
\includegraphics[width=\columnwidth]{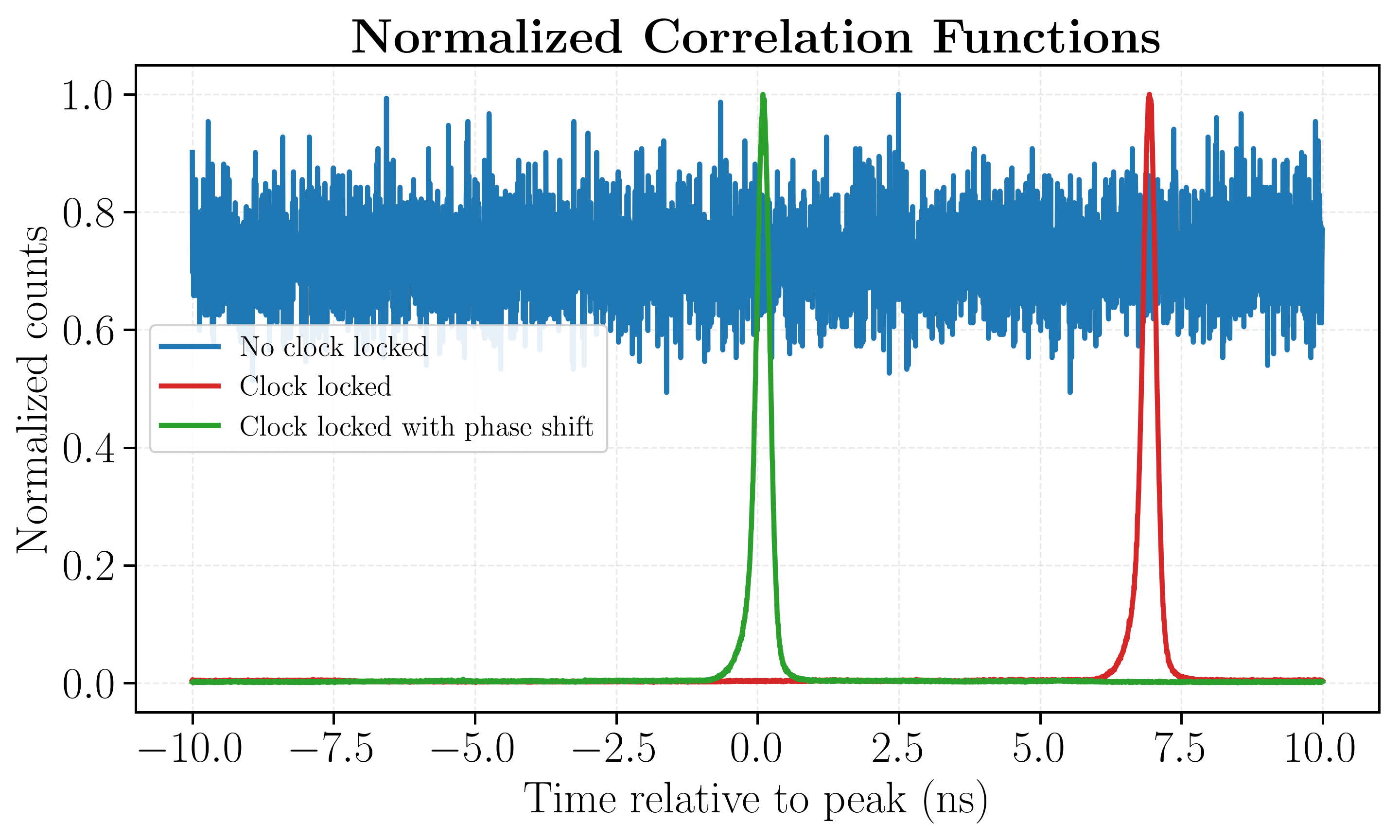}
\caption{Correlation histograms between detection channels CH1 and CH2 for different clock synchronization conditions of the two entangled photon sources. When the sources operate with identical clock periods but no clock sharing, no temporal correlations are observed. When the clock signal is electrically shared between the sources (clock lock condition), a clear coincidence peak emerges, demonstrating successful synchronization. Finally, by applying a controlled phase shift to one of the clocks, the correlation peak is displaced by the desired temporal offset (6.9 ns), enabling fine adjustment of the relative timing between the two sources.
 }
\hrulefill

\label{fig:03}
\end{figure}

In order to synchronize the two EPSs, they are configured in the same operating conditions, i.e. the clock period is set to 40 ns.
To determine whether the two sources are effectively synchronized to the same clock, the temporal correlation function between the photons generated by the two sources, $EPS_1$ and $EPS_2$, is analyzed. 
In particular, to compute the correlation function, the time tagger records the arrival time of every detection event on the two selected channels. Events associated with $EPS_1$ are taken as the temporal reference (\textit{start}), while those from $EPS_2$ are considered the corresponding \textit{stop} events.

For each detection pair, the time tagger computes the relative arrival-time difference and accumulates these values into a histogram over a time window equal to one period of the shared reference clock. If the two sources are synchronized, the photon pairs are detected with nearly the same relative delay, resulting in a pronounced peak in the histogram. Conversely, uncorrelated detection events are randomly distributed over the entire time window and contribute only to a nearly uniform background. Therefore, the resulting correlation function directly quantifies the temporal correlation between the two sources, while the position of the correlation peak provides the relative temporal offset between them.

For instance, we consider the first output of the beam splitter associated with $EPS_2$ and analyze the coincidences with photons from $EPS_1$ within a clock period. The presence of the peak thus constitutes initial experimental evidence of temporal coherence between the two sources and of their synchronization with respect to the shared clock.

However, we show that, setting the two sources at identical clock periods is not a sufficient condition to ensure synchronization between the entangled photon pairs. Indeed, no correlation is observed between the two outputs, as shown in Figure~\ref{fig:03}: only uncorrelated noise is present. This indicates that the two signals do not naturally synchronize or “lock” to each other, even if they have the same period\footnote{For the sake of clarity, the figure only shows the correlation between CH1 and CH2; however, the same considerations are valid and verifiable for all channel pairs in the proposed setup.}. Indeed, sharing the same clock period only ensures that the two EPSs run at the same frequency, but it does not guarantee temporal alignment, as an unknown relative time offset between them may still exist and must be estimated and compensated for.

To achieve synchronization, the clock signal has to be explicitly shared between the sources. In our setup, the rack-mounted source provides an accessible electrical clock output, enabling direct electrical distribution of the timing signal to the second source. Once the electrical clock is shared, a distinct peak appears in the correlation histogram, clearly indicating the presence of temporal correlations between the detected photon pairs.

Although the fiber lengths connecting the detectors are nominally identical, a temporal offset is eventually observed. As shown in Figure~\ref{fig:03}, the correlation peak between the two photons is shifted by approximately 6.9 ns. This residual delay is compensated by applying an appropriate control command to adjust the relative phase of the clock signals. By introducing a controlled shift in the clock of one source with respect to the other, temporal alignment is restored.

\begin{figure*}[!t]
\centering
\includegraphics[width=\textwidth]{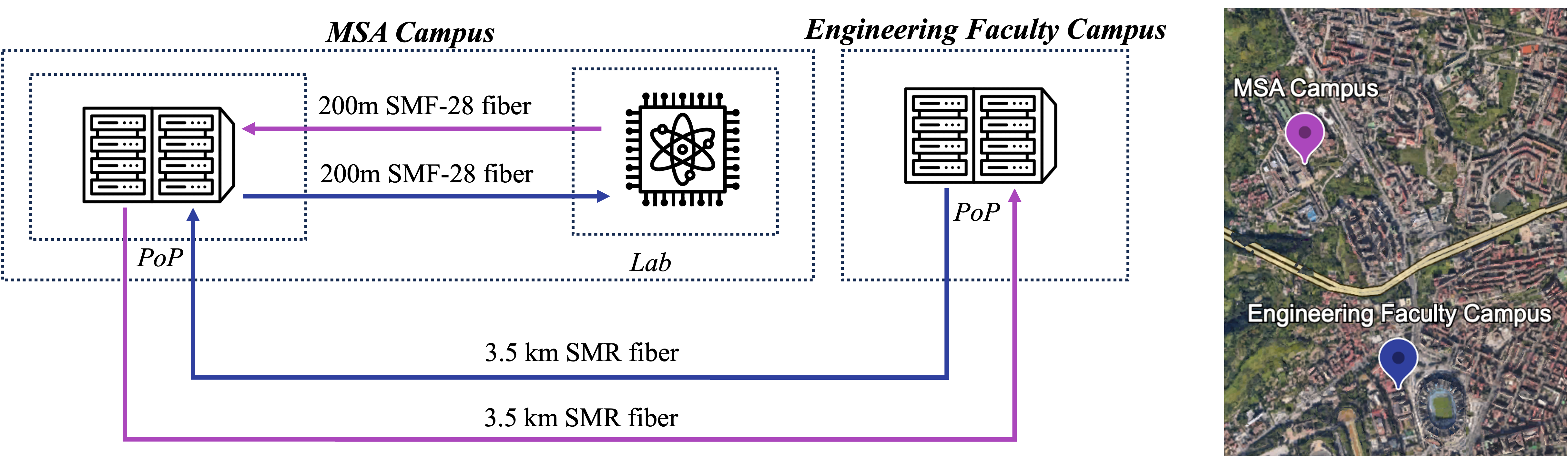}
\caption{Schematic of the 7 km fiber loop connecting the Monte Sant’Angelo campus (MSA) and the Engineering faculty campus. The entangled photon pairs are generated and detected in the laboratory at MSA and injected into the network at the local Point of Presence (PoP). The remote site is located approximately 3.5 km away, and the loop-back configuration results in a total propagation length of about 7 km.}
\hrulefill
\label{fig:loop}
\end{figure*}

\section{Metropolitan loop}
\label{sec:05}

Once the stability of the clock signal has been verified, the next goal is to distribute entanglement in a realistic metropolitan network scenario. However, the transition from a controlled laboratory environment to a metropolitan-scale implementation introduces significant challenges related to losses, noise, and imperfections in the optical infrastructure.
The experimental system is based on an optical loop connecting two university campuses: the MSA campus and a remote site located at the Faculty of Engineering, approximately 3.5 km away \cite{CalDavFlo-26}.  
Specifically, entangled photon pairs are generated and detected in the laboratory at the MSA campus and injected into the network through the local Point of Presence (PoP) at the same campus. The PoP is connected to the laboratory via approximately 200 m of SMF-28 fiber.

At the remote PoP located at the Faculty of Engineering Campus, the signal is looped back and sent back toward the MSA campus. The loop-back configuration results in a total propagation length of approximately 7 km, as shown in Figure~\ref{fig:loop}. 
This architecture allows for the study of the correlation between photons from the two sources in the presence of an actual installed fiber, thereby including effects due to imperfections in the telecommunications infrastructure.
In particular, the experiment is conducted on an existing metropolitan network that has been operational for over ten years. This implies the presence of non-ideal elements along the link, such as splices, connectors, and optical adapters (e.g., FC, LC, and SC interfaces), which contribute to the overall losses. The fibers used in the loop are standard single-mode reinforced (SMR) fibers, and the measured end-to-end loss is on the order of approximately 14 dB.

To counteract this attenuation in the loop, $EPS_2$ has a different configuration than the first one.
Specifically, the attenuation of the pump laser used for entanglement generation has been reduced compared to the one of $EPS_1$.
This choice is not determined solely by the fact that the generated photons are distributed across four distinct branches via beam splitters, but is primarily motivated by the significant losses introduced by the optical loop and the presence of non-negligible background noise, which requires compensation in terms of optical power to ensure an adequate signal-to-noise ratio in the measurements.
A crucial aspect concerns the characterization of the background noise present in the loop, namely, the presence of photons in the band of interest that are not associated with the injected signal but are generated by the infrastructure itself. To evaluate this contribution, the system is initially turned off, and the output signal from the loop is analyzed in the absence of optical input, measuring the detector count rate (cps) per channel. This procedure allows for the identification of any cross-talk effects or spurious contributions intrinsic to the system, independent of the transmitted signal.
To perform a more accurate spectral characterization of the noise, the loop output in the laboratory is subsequently connected to a cascade of DWDM filters, in order to isolate the individual sub-bands of the C-band. Downstream of the coarse selection, a tunable narrowband filter allows for a fine scan of the channels of interest prior to detection using a cryogenic system. This configuration thus enables detailed mapping of noise as a function of wavelength and a precise assessment of the background contributions present in the various C-band channels.
The results obtained confirm the importance of this preliminary analysis of background noise, highlighting the presence of significant contributions related to the actual optical infrastructure. In particular, a background level on the order of $10^3$ cps is observed, likely attributable to crosstalk effects and spurious emissions generated along the fibers and link interconnections \cite{CalDavFlo-26}.
Within this data set, a distinct peak emerges at channel 21, characterized by an increase of approximately two orders of magnitude compared to the average level. This behavior is attributable to the fact that this channel is dedicated to the University network traffic.
Further anomalies are also observed in the 46–61 channel region, where even more pronounced peaks are evident, with intensities up to three orders of magnitude higher than the background. The origin of these signals cannot be immediately traced to a single source and therefore requires further analysis. These contributions could be associated with combined effects of cross-talk between DWDM channels, multiple reflections, or spurious emissions introduced by active network components.
Overall, these observations underscore the need for a detailed spectral characterization of the infrastructure in order to reliably distinguish between the useful quantum signal and noise contributions induced by the actual metropolitan network.
\begin{figure}[!t]
\centerline{\includegraphics[width=\columnwidth]{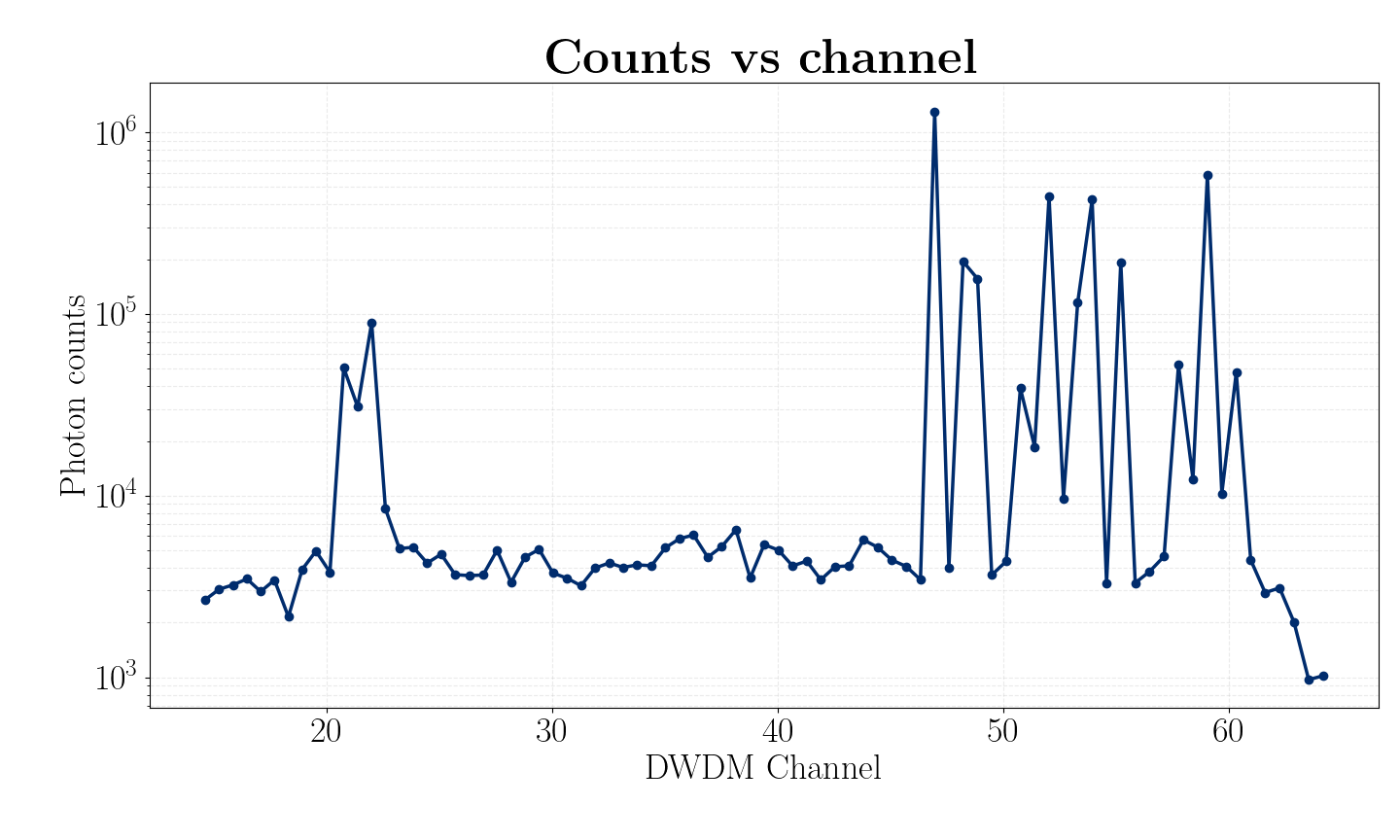}}
\caption{Measured noise count rate across DWDM channels of the deployed metropolitan loop. A background level of the order of $10^3$ cps is observed, with pronounced peaks around channel 21 (associated with the University network traffic) and in the range 46–61, where excess counts up to three orders of magnitude above the background are detected.}
\hrulefill

\label{fig:loop_noise}
\end{figure}
\begin{figure}[!t]
\includegraphics[width=\columnwidth]{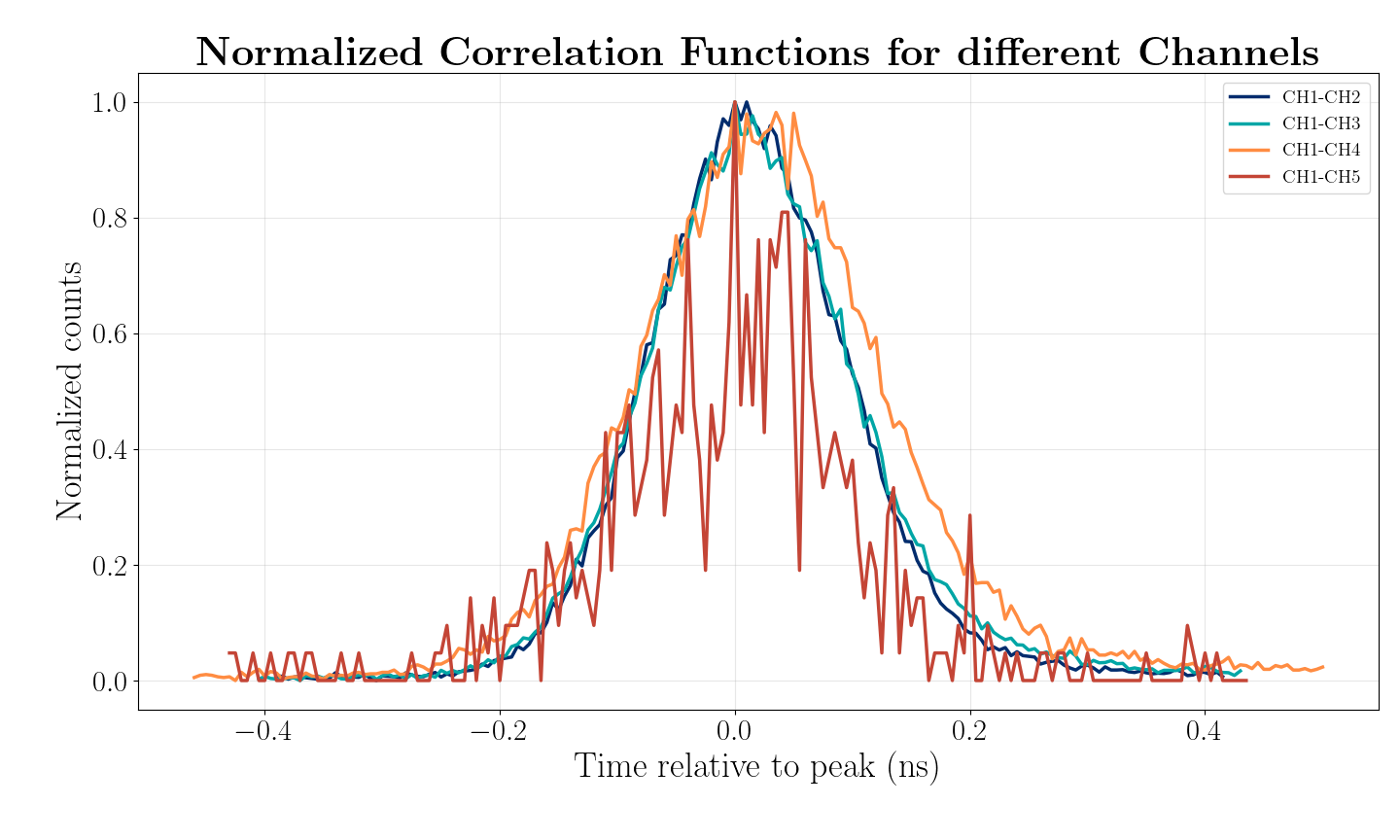}
\caption{Correlation distributions for different channel pairs over the first 5 minutes of measurement. While the metropolitan loop case exhibits noisier and more fluctuating profiles, all fits remain approximately Gaussian, indicating preserved photon correlation.}
    \hrulefill

\label{fig:09}
\end{figure}
In this regard, it is also important to note the decision to use channels 25-43, which are the most stable and least affected by spectral noise within the C-band. This choice effectively places the operation outside the noisy spectral regions identified in the previous analysis, thereby avoiding the channels affected by significant background spikes associated with the active network. As a result, the selected band provides more reliable conditions for correlation measurements and entanglement distribution.

Figure~\ref{fig:09}  shows a comparison of the correlation functions for different channel configurations\footnote{The Gaussian-fits shown refer to the first 5 minutes of measurement.}. 
In particular, in Figure~\ref{fig:09}, the correlation functions have been shifted by appropriate time delays in order to center them at zero. These delays depend on the length of the fibers and the specific operating conditions of the various channels. Furthermore, the counts have been normalized for clarity in the visualization, since the losses introduced by propagation over distances on the order of kilometers naturally result in a reduction in the number of detected counts.
In particular, the distribution obtained for the metropolitan loop appears more jagged and characterized by greater variability compared to the laboratory cases, indicating a more pronounced temporal dispersion. This aspect is further confirmed by the dispersion analysis, shown in Section~\ref{sec:Results} in Figure~\ref{fig:04b}, where an increase in the variability associated with the Gaussian-fit is observed for the metropolitan channel.
Nevertheless, it is important to emphasize that the overall trend remains well approximated by a Gaussian-fit. 
Therefore, despite the significant attenuation introduced by the metropolitan loop and the presence of a non-negligible level of background noise, the results demonstrate that, through proper optimization of the parameters of the two sources, it is still possible to effectively distribute entangled traffic within a real metropolitan environment. This result represents a first concrete step toward the distribution of entanglement in real-world scenarios by leveraging existing and operational optical infrastructure. In this sense, the work highlights the possibility of integrating entanglement distribution into conventional telecommunications networks. It is important to note, however, that the counts used to derive the correlation function, in the case of the metropolitan loop, exhibit a distribution that differs from that obtained in a controlled laboratory environment. In the latter case, measurements are taken under stable conditions -- for example, with the temperature kept constant (around $24\degree$ C) throughout the experiment -- thereby minimizing external fluctuations. In the metropolitan context, however, while maintaining an approximately Gaussian trend, the distribution is less regular and more affected by fluctuations, not merely because we are outside a controlled environment, but precisely due to the contribution of noise and channel imperfections. This behavior is directly reflected in the correlation function between photons distributed across distinct channels.

\section{Results}
\label{sec:Results}

\begin{figure*}[htbp]
\centering

\begin{subfigure}{0.49\textwidth}
    \centering
    \includegraphics[width=\linewidth]{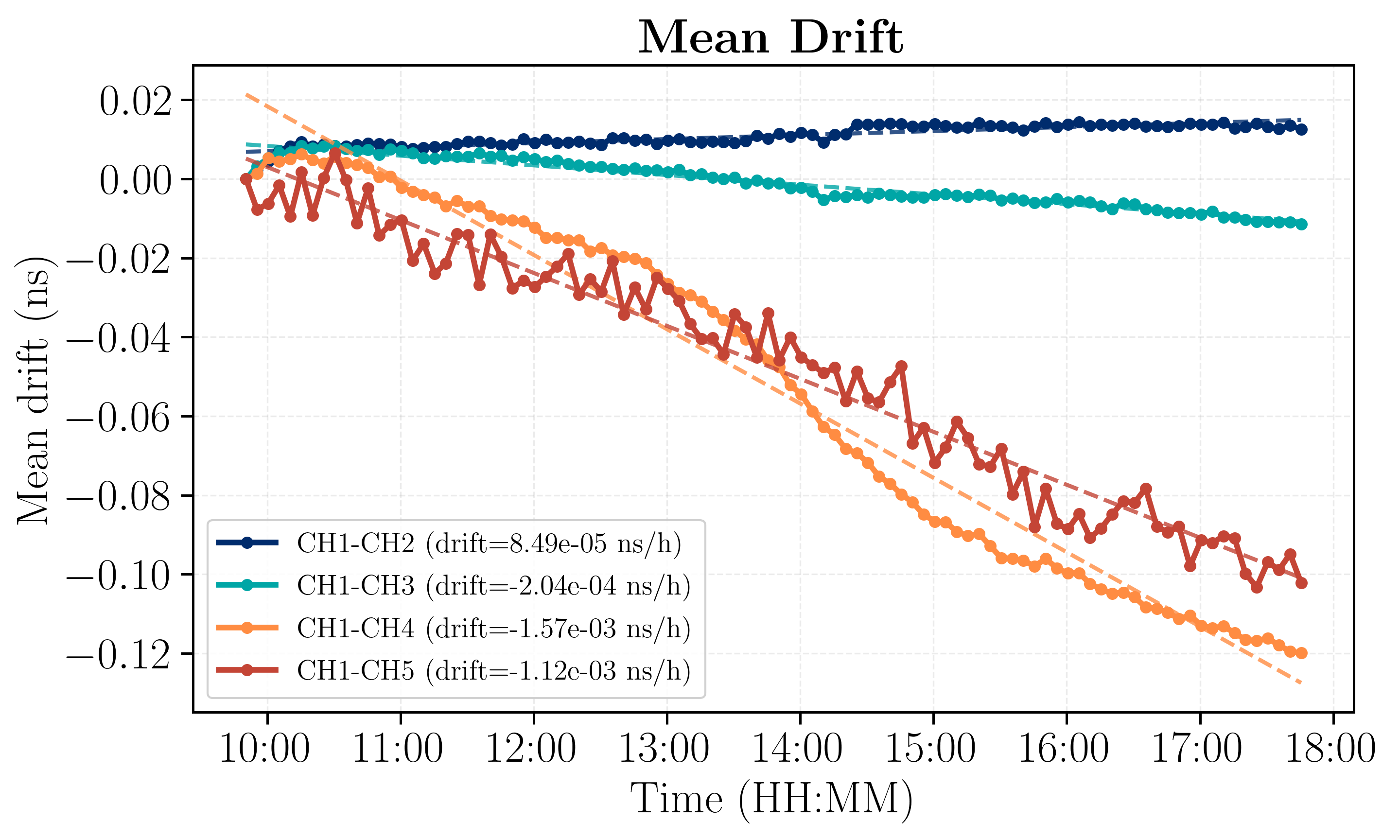}
    \caption{Temporal evolution of the mean value of the Gaussian-fit used to approximate the correlation function for different channel pairs. Linear fits (dashed lines) are shown to highlight the presence of a possible time drift, while the corresponding slopes provide a quantitative estimate of the drift rate for each pair of channels.}
    \label{fig:04a}
\end{subfigure}
\hfill
\begin{subfigure}{0.49\textwidth}
    \centering
    \includegraphics[width=\linewidth]{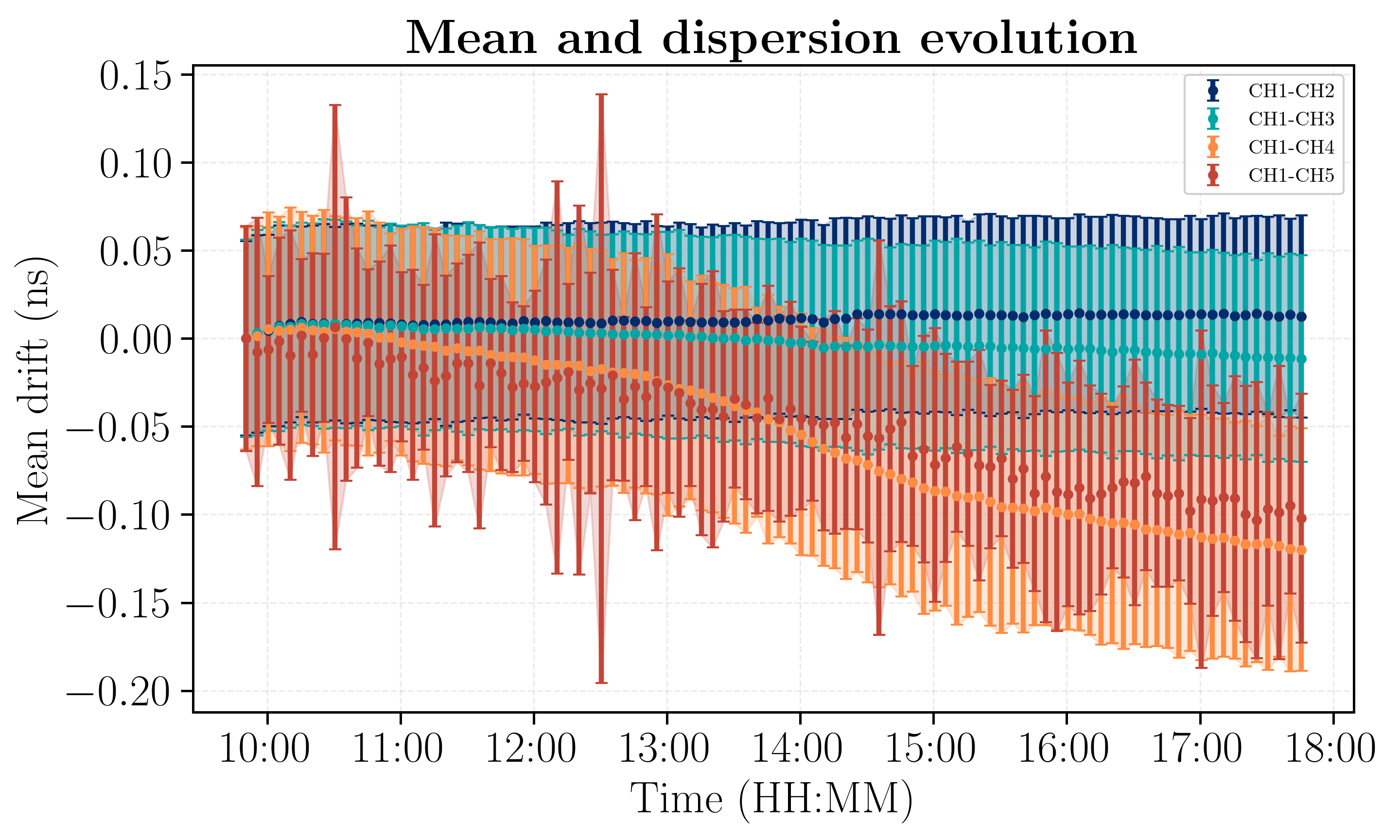}
    \caption{Temporal evolution of the dispersion of the correlation mean, evaluated as the interquartile range (P75–P25). The results show that in the laboratory case the temporal width of the correlation peak remains approximately constant ($\approx100$ ps), indicating stable system performance over time. In the metropolitan loop scenario, instead, a larger variability is observed, likely due to increased environmental noise and fluctuations in the optical infrastructure.}

    \label{fig:04b}
\end{subfigure}

\caption{Temporal evolution of the mean correlation peak and its dispersion across multiple channel pairs over an 8-hour measurement interval.}
\hrulefill
\label{fig:04}
\end{figure*}

\begin{table*}[t]
\centering
\renewcommand{\arraystretch}{1.2}
\setlength{\tabcolsep}{6pt}
\caption{Temporal synchronization parameters for different channel pairs}
\label{tab:01}

\small

\begin{tabular}{|c|c|c|c|c|}
\hline\hline

 & \textbf{CH1-CH2} & \textbf{CH1-CH3} & \textbf{CH1-CH4} & \textbf{CH1-CH5} \\
\hline\hline

\textbf{Channel type} 
& Reference link & 500 m spool & 5 km spool & 7 km metropolitan loop \\
\hline

\textbf{Initial time shift (ps)} 
& +4500 & -5970 & +2710 & +6320 \\
\hline

\textbf{Measured drift in 8 hours (ps)} 
& +12.56 & -11.46 & -119.91 & -102.16 \\
\hline

\textbf{Initial width (p75-p25) (ps)} 
& 110.38 & 112.02 & 126.07 & 127.72 \\
\hline

\textbf{Mean width variation (ps)} 
& +4.33 & +5.24 & +11.64 & +13.52 \\
\hline\hline

\end{tabular}
\end{table*}

\begin{figure*}[!t]
\centering

\begin{subfigure}{0.49\textwidth}
    \centering
    \includegraphics[width=\linewidth]{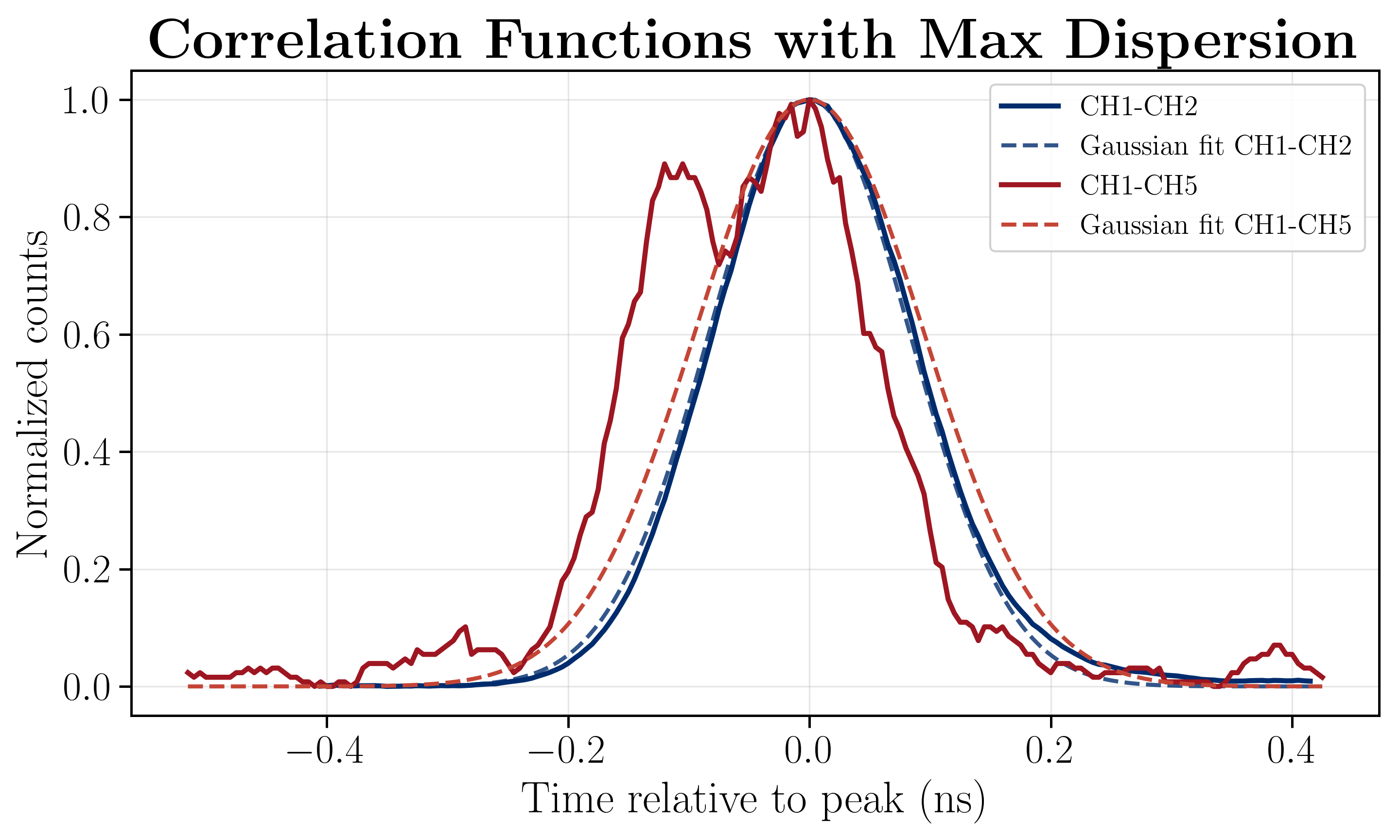}
    \caption{Worst-case scenario: comparison between CH1-CH2 and CH1-CH5 correlation histograms.}
    \label{fig:worst}
\end{subfigure}
\hfill
\begin{subfigure}{0.49\textwidth}
    \centering
    \includegraphics[width=\linewidth]{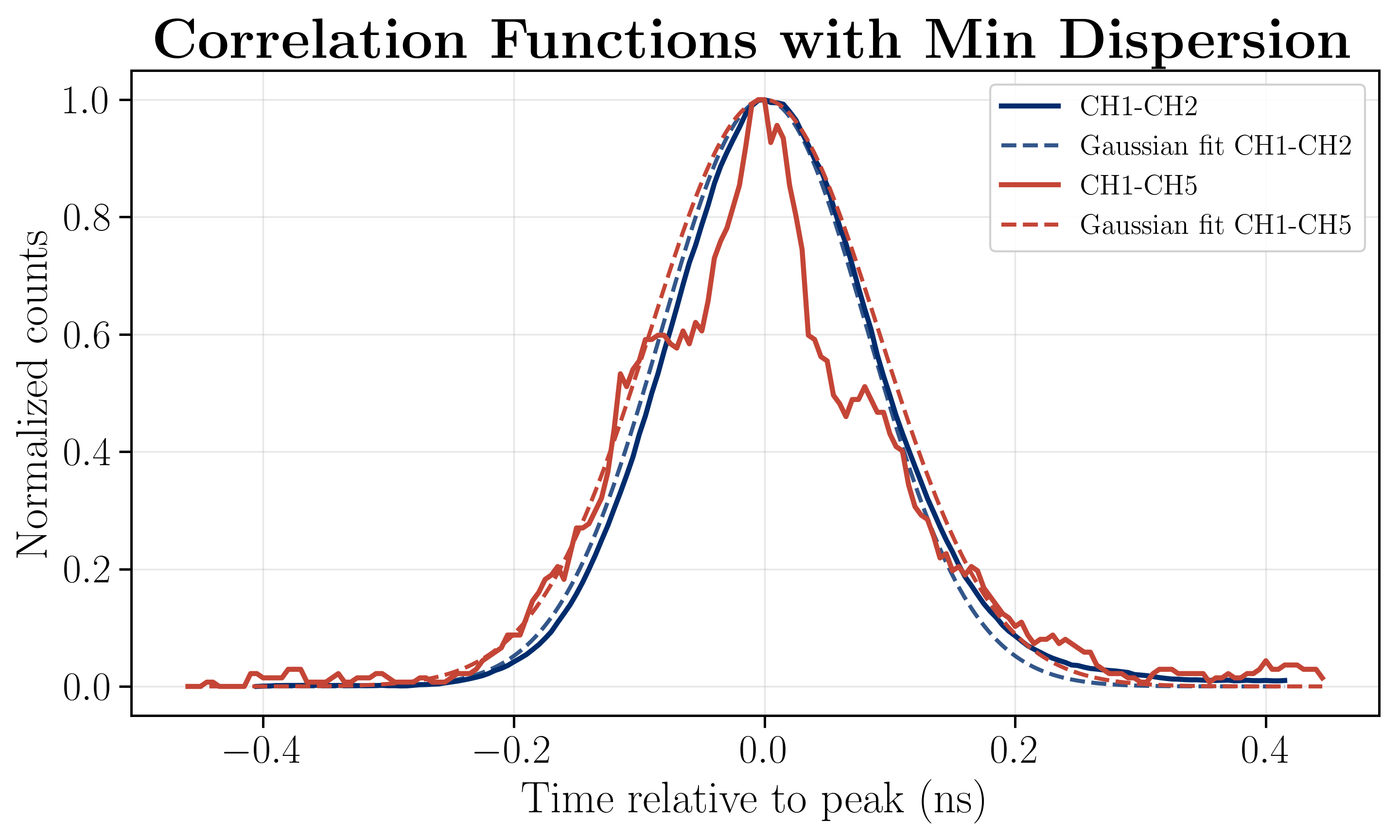}
    \caption{Best-case scenario: comparison between CH1-CH2 and CH1-CH5 correlation histograms.}
    \label{fig:best}
\end{subfigure}

\caption{Temporal correlation analysis for the CH1-CH2 (laboratory reference link) and CH1-CH5 (metropolitan loop) links under worst-case (largest observed temporal fluctuations) and best-case (smallest observed temporal fluctuations) conditions, together with the corresponding Gaussian-fits.}
\hrulefill
\label{fig:nuova}
\end{figure*}

\begin{figure}[!t]
\includegraphics[width=\columnwidth]{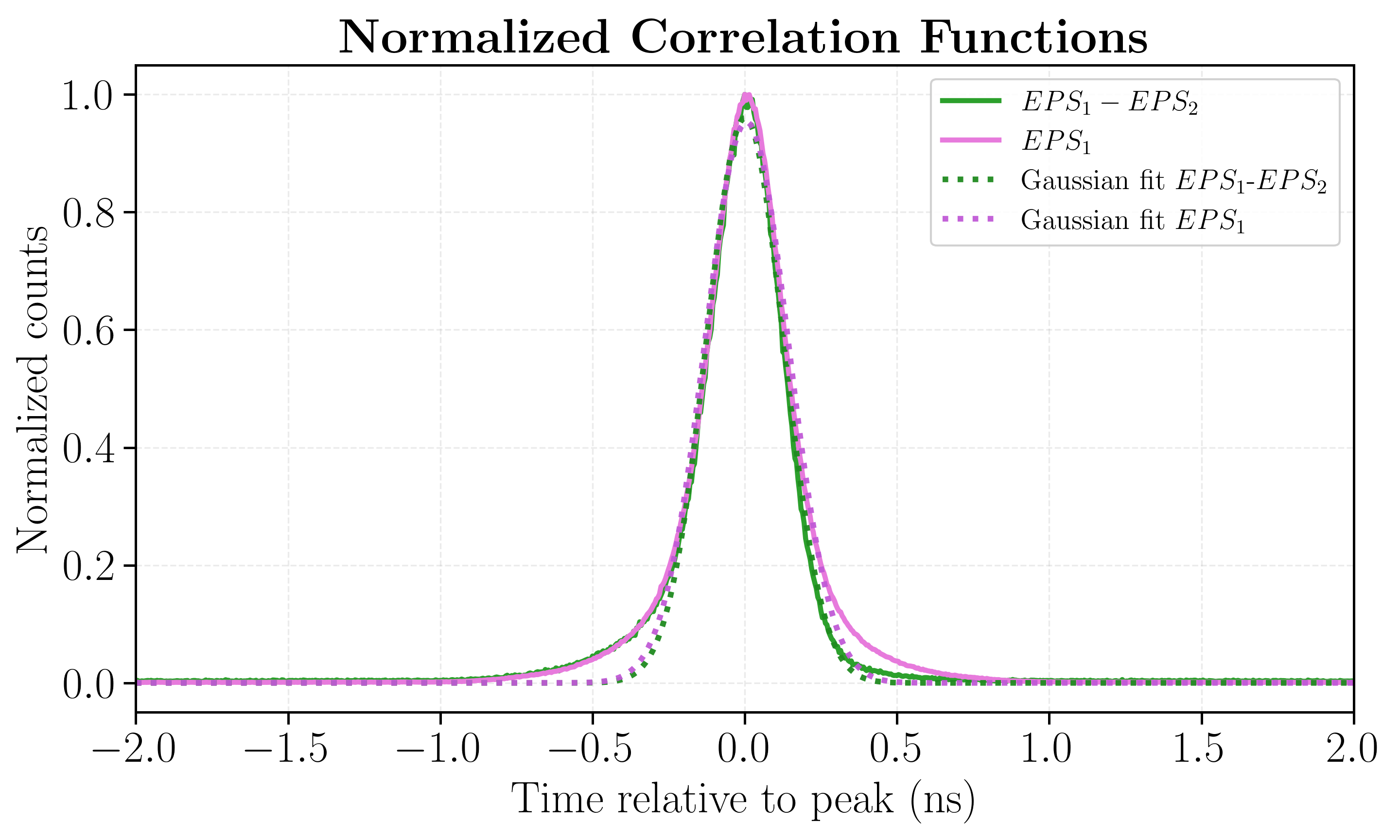}
\caption{Comparison between the correlation functions of photon counts generated from $EPS_1$ (DWDM channels 43–25) and those obtained from the two sources $EPS_1$ and $EPS_2$ (DWDM channels 43–43).}
\hrulefill
\label{fig:06}
\end{figure}

Once we have established that the clock can be shared among two sources and correlation can be distributed in a metropolitan-scale fiber infrastructure, we verify the clock stability. This is particularly important when the clock is shared among sources that distribute entanglement on a large scale via optical fibers of varying lengths. In this context, it is essential to ensure that the correlation remains temporally stable—that is, that it does not drift over time.
To assess long-term stability, we perform a continuous 8-hour measurement, with data collected every 5 minutes. The resulting correlation function can be approximated by a Gaussian-fit\footnote{The resulting correlation function can be approximated by a Gaussian-fit, as evidenced in Figure~\ref{fig:03}  and Figure~\ref{fig:09}. A Gaussian shape is also visible later in Figures~\ref{fig:nuova} and Figure~\ref{fig:06}.}.

Our study therefore focuses on two main aspects, (i) the temporal stability of the Gaussian-fit mean and (ii) the dispersion of the Gaussian-fit over time. Experimental results are shown in Figure~\ref{fig:04}.

Regarding the first aspect, we analyze any variation in the mean value of the Gaussian-fit over time, i.e., the eventual shift in the peak of the correlations. In the presence of systematic drift, it is possible to intervene by recalibrating the clock to compensate for this variation. Figure~\ref{fig:04a} shows the measurement results, where both the mean values of the Gaussian-fits (evaluated for each 5-minute acquisition) and the corresponding linear fit, whose slope quantifies the clock drift.
An initial drift is observed within the first 10 minutes of measurement, which is attributed to the transient settling of the clock synchronization system between the two devices. In this initial phase, the system requires a finite time to reach a stable synchronization before converging to a steady-state condition in which the measurements are reliable and reproducible.

The results show that a significant variation in the clock occurs exclusively in the channels associated with longer fibers, particularly in the 5-km spools and the 7-km loop. The maximum drift is observed after approximately 8 hours of acquisition, reaching a value of 0.12 ns (120 ps) in CH1-CH4 correlation measurement on the 5 km fiber spool. 

This behavior indicates that, beyond this time interval, it becomes necessary to adjust the clock of the $EPS_2$ source so that it can realign with the reference clock provided by $EPS_1$, in order to maintain synchronization between the two systems. Furthermore, this result is consistent with the maximum correction limit (clock shifting) applicable via $EPS_2$, which in our case is on the order of approximately 10 ns, thus far exceeding the observed drift.

Regarding the second aspect, we analyze the dispersion of the Gaussian-fit over time. Any variation in this quantity would indicate a deterioration in the temporal precision of the correlations between entangled photons and, consequently, in the overall performance of the synchronization system.
In particular, a variation of approximately 100 ps is observed, estimated as the interquartile range (between P25 and P75) of the distribution\footnote{P25 and P75 denote the 25th and 75th percentiles of the fitted distribution, respectively. Their difference defines the interquartile range (IQR), which quantifies the spread of the central 50\% of the distribution. The IQR is widely used as a robust estimator of statistical dispersion because it is less affected by outliers than the full width of the distribution; for approximately Gaussian distributions, it provides a reliable measure of timing jitter.}. This value is comparable to the intrinsic jitter of the single-photon detectors used ($\approx100$ ps) \cite{NatTanHad-12}. Consequently, the measured correlation width is likely instrument-limited, indicating that it is predominantly determined by the detector timing resolution rather than by any significant broadening arising from the underlying physical processes.
In addition to the average trend, we perform a worst-case analysis by identifying the time intervals during which the greatest temporal fluctuations occur. In particular, we focus on the metropolitan loop configuration, which appears to be the most affected in terms of increased dispersion.
It is observed that the maximum dispersion of the CH1–CH5 correlation function in the metropolitan loop configuration occurs approximately at 12:30 PM, when environmental fluctuations cause the most significant widening of the Gaussian-fit. In contrast, the CH1–CH2 correlation shows no significant fluctuations, and its maximum dispersion is recorded around 11:20 AM.
To provide a comparison, Figure~\ref{fig:nuova} shows the correlation functions for these worst-case scenarios, along with the respective Gaussian-fits used as a reference. 
The time axis is limited to a window equal to a portion of the total period under consideration (40 ns), and the counts have been normalized to facilitate visualization and comparison between the cases.
This comparison clearly highlights the increase in temporal dispersion compared to nominal conditions. For completeness, a representative example of a low-dispersion regime is also shown, observed around 10:00 AM for CH1-CH5 and at 11:55 AM for CH1-CH2, in which the correlation function exhibits a well-defined and narrow Gaussian-fit, consistent with optimal synchronization conditions.

Overall, this comparison shows that, while the system maintains stable synchronization under nominal conditions, localized temporal fluctuations -- particularly in metropolitan settings -- can cause a temporary increase in the correlation width. The greater variability is observed in the case of the metropolitan loop compared to laboratory conditions, an effect attributable primarily to additional noise and the uncontrolled nature of the metropolitan environment.
However, even in the worst-case scenarios considered, the dispersion remains within the range limited by the detectors’ resolution, confirming the robustness of the synchronization throughout the entire 8-hour acquisition interval. 

Finally, Table~\ref{tab:01} summarizes both the main characteristics of the different channels and relevant quantitative values extracted from the analysis.
We report the initial time offset, the long-term temporal drift, and the evolution of the correlation width, quantified as the interquartile range (P75-P25).
Short fiber links exhibit stable temporal widths of approximately 110 ps, with minimal variation over time ($<6$ ps), indicating a high degree of temporal stability of both sources. In contrast, longer transmission paths (5 km and 7 km) show both increased initial temporal dispersion and a more pronounced broadening over time, reaching up to $\approx13$ ps. This behavior suggests an enhanced sensitivity to environmental fluctuations and propagation-induced instabilities in extended fiber links.

Furthermore, we aim to verify whether the dispersion of the Gaussian-fit is influenced by the fact that the photons are generated by different sources. To this end, we measure the correlation function of the photon counts from DWDM channels 43–25 of the $EPS_1$, where the fiber length of channel 25 is chosen to match the fiber length of channel 43.
Figure~\ref{fig:06} shows both correlation histograms: one for photons generated by the same source, $EPS_1$, and one obtained from photons originating from different sources. Specifically, we chose to show in Figure~\ref{fig:06} CH1–CH2 correlation histogram, in order to maintain approximately the same optical channel length in the comparison\footnote{Similarly to the previous correlation histograms, also in this case the time axis has been narrowed, and the counts have been normalized to facilitate comparison between the distributions. However, in Figure~\ref{fig:06} the Gaussian-fit is narrower and concentrated within a shorter time interval, since the measurements were taken under operating conditions of the two sources that differ from those adopted in the proposed configuration of our setup, shown in Figure~\ref{fig:02}. Indeed, setting the sources in our setup requires a significant imbalance of some parameters, as will be clarified in Section \ref{sec:05}.}.
In the case of photons from the same source, the correlation function exhibits a symmetric and well-defined profile. In contrast, when photons come from different sources, a slight asymmetry in the correlation function is observed. However, this effect is not accompanied by a significant change in the amplitude or width of the distribution. This result suggests that the dispersion remains essentially unchanged, while differences between the sources manifest primarily through small distortions in the symmetry of the correlation function.

\section{Discussion}

Overall, the results show that the position of the correlation peak remains essentially stable over time, with no significant fluctuations. This confirms that the distributed clock provides a robust synchronization backbone, capable of sustaining long-duration operation without the need for active real-time feedback or continuous recalibration.
The slight drift observed over longer time scales is primarily attributable to propagation in longer optical fibers. In particular, environmental variations, especially in temperature, induce changes in both the refractive index and the effective optical length of the fiber (thermo-optical effects and thermal expansion). These phenomena result in variations in signal propagation time, which become more pronounced as the fiber length increases. The measured drift is therefore consistent with this physical mechanism and is compatible with the time scales and lengths considered. In addition, the dispersion of the Gaussian-fit remains essentially invariant over time, indicating that no significant temporal broadening is introduced. The measured correlation width is instead consistent with the intrinsic timing jitter of the SNSPDs, confirming that the system operates in a detector-limited regime.
The observation of stable temporal correlations between two distant sources, despite the significant attenuation caused by varying fiber lengths, the metropolitan loop, and the presence of significant background noise, constitutes a fundamental result. It experimentally demonstrates the possibility of preserving the correlation structure of entanglement even under realistic operating conditions typical of non-dedicated metropolitan infrastructures.

\bibliographystyle{IEEEtran}
\bibliography{bibliography.bib}

\end{document}